\documentclass[apj]{emulateapj}
\usepackage{amsmath}                
\usepackage{amsfonts}               
\usepackage{amssymb}                
\usepackage{epsfig}                 
\usepackage[para,online,flushleft]{threeparttable}

\newcommand{\km}{{~\rm km}}
\newcommand{\s}{{~\rm s}}

\newcommand{\erg}{{~\rm erg}}

\newcommand{\nar}{{~\rm New Astronomy Reviews}}

\begin{document}

\title{Reviving the stalled shock by jittering jets in core collapse supernovae: jets from the standing accretion shock instability}

\author{Noam Soker\altaffilmark{1,2}}

\altaffiltext{1}{Department of Physics, Technion -- Israel Institute of Technology, Haifa
32000, Israel; soker@physics.technion.ac.il}
\altaffiltext{2}{Guangdong Technion Israel Institute of Technology, Shantou 515069, Guangdong Province, China}

\begin{abstract}
I present a scenario by which an accretion flow with alternating angular momentum on to a newly born neutron star in core collapse supernovae (CCSNe) efficiently amplifies magnetic fields and by that launches jets. The accretion flow of a collapsing core on to the newly born neutron star suffers the spiral standing accretion shock instability (SASI). This instability leads to a stochastically variable angular momentum of the accreted gas, that in turn forms an accretion flow with alternating directions of the angular momentum, hence alternating shear, at any given time.
I study the shear in this alternating-shear sub-Keplerian inflow in published simulations, and present a new comparison with Keplerian accretion disks. From that comparison I argue that it might be as efficient as Keplerian accretion disks in amplifying magnetic fields by a dynamo. I suggest that although the average specific angular momentum of the accretion flow is small, namely, sub-Keplerian, this alternating-shear accretion flow can launch jets with varying directions, namely, jittering jets. Neutrino heating is an important ingredients in further energizing the jets. The jittering jets locally revive the stalled accretion shock in the momentarily polar directions, and by that they explode the star. I repeat again my call for a \textit{paradigm shift from a neutrino-driven explosion of CCSNe to a jet-driven explosion mechanism that is aided by neutrino heating.}
\end{abstract}

\section{Introduction}
\label{sec:intro}

Simulations of collapsing massive stars over the years did not reach a consistent and robust explosion in the frame of the delayed neutrino mechanism. Two recent examples of contradicting results are the claim made by \cite{Mulleretal2017} for a successful explosion of a core collapse supernova (CCSN) versus the finding by \cite{OConnorCouch2018} of no explosion. In a third recent paper, \cite{Vartanyanetal2019} manage to exploded the inner part of the core, but they do not reach a positive total energy, so they do reach a successful 3D CCSN explosion model.
It seems that the delayed neutrino mechanism has generic problems (e.g., \citealt{Papishetal2015, Kushnir2015b}).

In light of these difficulties of the thirty four years old delayed neutrino mechanism we have developed the jittering-jets explosion mechanism, that we suggest explodes all or most CCSNe (e.g., \citealt{PapishSoker2011, GilkisSoker2015}).
It seems that neutrino heating does play a role in the jittering jets explosion mechanism by keeping the outflowing gas hot, and by that supplying more  energy to the bipolar outflow \citep{Soker2019}.
We then extended the model to include super-energetic (or super luminous) CCSNe that are exploded by
the more general jet feedback mechanism (\citealt{Gilkisetal2016, Soker2017RAA}; for a review see \citealt{Soker2016Rev}). There is mounting observational evidence from  the morphological features of some supernova remnants and from polarizations of some CCSNe that jets play a role in many, and possibly in most, CCSNe (e.g., \citealt{Wangetal2001, Maundetal2007, Lopezetal2011, Milisavljevic2013, Gonzalezetal2014, Marguttietal2014, Inserraetal2016, Mauerhanetal2017, GrichenerSoker2017, Bearetal2017, Garciaetal2017,  LopezFesen2018}).

It is important to emphasise the unique characteristics of the jittering jets explosion mechanism.
There were many studies of jet-driven explosion of massive stars before and after the development of the jittering-jets explosion mechanism. However, these studies were aiming at particular CCSNe, those that the core of their progenitors was rapidly rotating before explosion, and hence the jets maintain a constant direction (e.g., \citealt{Khokhlovetal1999, Aloyetal2000, Hoflich2001, MacFadyen2001, Obergaulingeretal2006, Burrows2007, Nagakuraetal2011, TakiwakiKotake2011, Lazzati2012, Maedaetal2012, LopezCamaraetal2013, Mostaetal2014, LopezCamaraetal2014, Itoetal2015, BrombergTchekhovskoy2016, LopezCamaraetal2016, Nishimuraetal2017, Fengetal2018, Gilkis2018}). These studies consider jet-driven explosions to be rare because a stellar binary companion must spin-up the pre-collapse core (at least in metal-rich stars).

The jittering-jets explosion mechanism has these unique properties. (1) It explodes all CCSNe, at least those with kinetic energies of $\ga 10^{50}\erg$. (2) The pre-collapse core needs not be rapidly rotating. (3) The jets might have varying directions and can be intermittent. (4) The jets operate in a negative feedback mechanism. Namely, the jets reduce the accretion rate and hence their power while removing mass from the core. (5) Each pair of jets in the many jittering-jets that are launched in the explosion lives for a short time. Therefore, in general these jets do not break out from the exploding envelope and might leave only small imprints on the explosion and the remnant. In some cases the last jets to be launched might reach the outer boundary of the already expanding envelope and form two opposite small lobes (ears) in the supernova remnant (e.g., \citealt{GrichenerSoker2017, Bearetal2017}).

Some dynamical processes can ease the revival of the stalled shock in the delayed neutrino mechanism.
One such mechanism is the introduction of convection-driven perturbations (or turbulence) in the core of the massive star before collapse starts (e.g., \citealt{Mulleretal2017}), that in turn lead to fluctuations in the magnitude and direction of the specific angular momentum of the core mass that is accreted on to the newly born neutron star, or onto a black hole if accretion continues to include the helium and hydrogen zones of the star  \citep{GilkisSoker2014, GilkisSoker2015, Quataertetal2019}.
Instabilities, like the neutrino-heated bubbles that push the downflows around \citep{Mulleretal2017, Kazeronietal2018}, but mainly the spiral modes of the standing accretion shock instability (SASI), increase the angular momentum stochastic amplitudes when the gas reaches the neutron star vicinity.

The spiral-SASI modes (e.g., \citealt{BlondinMezzacappa2007, Rantsiouetal2011, Fernandez2010, Iwakamietal2014, Kurodaetal2014, Fernandez2015, Kazeronietal2017}) that develop between the shock of the inflowing gas at $\approx 100 \km$ and down to the neutron star at $\approx 20-40 \km$ form an accretion flow with a specific angular momentum that changes its sense at any given time. Namely, while some parcels of gas move clockwise, others in the vicinity moves counterclockwise, forming a general spiral structure when one draws the direction of the angular velocity in a plane that is perpendicular to the momentarily direction of the angular momentum. This flow has an alternating shear. As well, the angular momentum axis changes with time (e.g., \citealt{Hankeetal2013}).

Studies of the jittering-jets explosion mechanism have been assuming that the spiral SASI forms an accretion belt around the newly born neutron star, and the belt launches the jittering jets (e.g., \citealt{SchreierSoker2016, Soker2019}). In the accretion belt the gas orbits the accreting object very close to its surface, with a sub-Keplerian specific angular momentum. At any given time the gas in the accretion belt orbits the accreting body in the same direction.
Namely, the accretion belt scenario for launching jets considers only the very inner part of the spiral-SASI structure.
In particular, in a recent paper \citep{Soker2019} where I consider the accretion belt scenario I argue that numerical simulations must include magnetic fields if they are to explore the explosion mechanism of CCSNe.
Some studies (e.g., \citealt{Masadaetal2015, Mostaetal2015, ObergaulingerAloy2017, Obergaulingeretal2018}) have taken the first direction in exploring the role of magnetic fields by high resolution simulations.

In the present study I set the goal to study the entire volume of the spiral SASI, from the NS and out to the shock and compare it to Keplerian accretion disks. This region might be more likely to launch the jittering jets that explode CCSNe than an accretion belt that was studied in earlier papers of the jittering jets explosion mechanism.
As well, the launching of jets from a much larger region, and in particular from the gain region where neutrino heating is important, can make a more efficient use of energy that is supplied by neutrino heating.
In conducting this study I am motivated by the new results of \cite{OConnorCouch2018} that find no explosion in their core collapse simulations, but do find strong spiral-SASI modes.

\section{Comparing to Keplerian accretion disks}
\label{sec:alternating}

There are tens of different numerical simulations in 2D and 3D of the spiral-SASI
(e.g., \citealt{BlondinMezzacappa2007, BlondinShaw2007, Rantsiouetal2011, Fernandez2010, Hankeetal2013,  Iwakamietal2014, Kurodaetal2014, Fernandez2015,  Blondinetal2017, Kazeronietal2017}).
Most relevant are the simulations by \cite{Endeveetal2010} and \cite{Endeveetal2012} who study the amplification of magnetic fields by the spiral-SASI modes. \cite{Endeveetal2012} find that outside the neutrinosphere the SASI can substantially increase the strength of the magnetic fields and \cite{Endeveetal2010} find the amplification to be by about four orders of magnitude. Some other studies, on the other hand, find much smaller amplification factors of the magnetic field intensity (e.g., \citealt{Obergaulingeretal2009, Obergaulingeretal2014, Rembiaszetal2016a, Rembiaszetal2016b}).
However, numerical MHD simulations are highly limited by resolution, i.e., numerical resistivity suppresses the growth of the magnetic fields (e.g., \citealt{Endeveetal2010, Endeveetal2012}), and hence in these studies the results are limited. As well, they did not refer to the possibility of launching jets, and did not make a comparison to Keplerian accretion disks.

Below I do not calculate the amplification of the magnetic fields, but I rather limit myself to comparison with Keplerian accretion disks that we observationally know that are capable of launching jets.

 The relevant quantities for the comparison are the angular velocity $\Omega$ and the the shear, $d \Omega/dr$. In a thin Keplerian accretion disk these quantities are important only in the equatorial plane and they depend there only on the distance from the center $r$. In the spiral SASI the velocity ${\bf v}$ and angular velocity depend also on the azimuthal angle and hence we better use the vorticity $\omega=\nabla \times {\bf v}$. The growth rate and the equilibrium value of the magnetic fields increases as the shear (or vorticity) increases. In the $\alpha \Omega$ dynamo the growth rate of the field depends on $d \Omega /dr$. In the dynamo model of \cite{Spruit2002} for non-convective zones of stars, that \cite{SchreierSoker2016} used for their belt model, the strength of the equilibrium magnetic field depends on $q \equiv \sqrt {r \Omega d \Omega /dr}$.

The numerical values of two of these quantities for a Keplerian accretion disk are
\begin{equation}
\begin{split}
\omega_{\rm Kep} & = (\nabla \times {\bf v})_{\rm Kep}
=\frac{1}{2} \Omega_{\rm Kep}(r)
\\
&
=
560 \left( \frac{M_{\rm NS}}{1.2M_\odot} \right)^{1/2}   
\left( \frac{r}{50 \km} \right)^{3/2} s^{-1} ,
\end{split}
\label{eq:rotVK}
\end{equation}
and
\begin{equation}
\begin{split}
q_{\rm Kep} & = \left( r \Omega \frac{d \Omega}{d r} \right)_{\rm Kep}^{1/2}
\\
&
=
1400 \left( \frac{M_{\rm NS}}{1.2M_\odot} \right)^{1/2}   
\left( \frac{r}{50 \km} \right)^{3/2} s^{-1} .
\end{split}
\label{eq:qKep}
\end{equation}

\cite{Endeveetal2010} find the vorticity in the region $40 \km \la r \la 100 \km$ to be in the range of
$100 \s^{-1} \la (\nabla \times {\bf v})_{\rm SASI} \la 10^ 4 \s^{-1}$ (see also \citealt{Endeveetal2012}). Their neutron star mass is $1.2 M_\odot$ and we can compare it to the vorticity in a Keplerian accretion disk as given in equation (\ref{eq:rotVK}). This yields
\begin{equation}
0.2 \la \frac{(\nabla \times {\bf v})_{\rm SASI} } {(\nabla \times {\bf v})_{\rm Kep} } \la 20
\label{eq:rotVcompare}
\end{equation}
Over all, the vorticity in the 3D numerical simulation of \cite{Endeveetal2010} is larger than that in a Keplerian accretion disk.

I turn to the quantity $q$ as I infer from the recent 3D simulations of \cite{OConnorCouch2018} and the older simulations by \cite{Hankeetal2013} who have similar numbers.
From figure 8 of \cite{OConnorCouch2018} and figure 7 of \cite{Hankeetal2013} the typical rotational velocity in the SASI zone of $30 \km \la r \la 100 \km$ is $A_{\rm SASI} \simeq 2 \times 10^4 \km \s^{-1}$. In many small regions the velocity reaches twice as large values. For a neutron star mass of $\simeq 1.6  M_\odot$ the Keplerian velocity is $v_{\rm Kep} = 6 \times 10^4 (r/60 \km )^{-1/2} \km \s^{-1}$. From these I find $A_{\rm SASI} \approx (1/3) v_{\rm Kep}$.
In many regions inside the spiral SASI zone the variation of the velocity from this typical value to zero velocity occurs over a typical distance of $\Delta r \approx 0.1 r$. This gives a value of
\begin{equation}
q_{\rm SASI} \approx \left( r 0.3 \Omega_{\rm Kep} \frac{0.3 \Omega_{\rm Kep}}{0.1 r} \right)^{1/2} \approx q_{\rm Kep}
\label{eq:qcompare}
\end{equation}
over most of the SASI zone. In some regions the angular velocity changes from $\approx +3 \times 10^4 \km \s^{-1}$ to   $\approx -3 \times 10^4 \km \s^{-1}$  within few km. This gives small regions with $(\nabla \times {\bf v})_{\rm SASI} \approx 2 \times 10^4 \s^{-1}$, which is more than an order of magnitude larger than the value for a Keplerian accretion disk.

The conclusion from this section is that the spiral SASI modes that amplify pre-collapse perturbations lead to an accretion flow with shear and vorticity that are comparable to those in Keplerian accretion disks. The relevant point to this study is that Keplerian disks are known to be capable to launch jets. From that I speculate that the spiral-SASI can also launch jets.

There is one caveat to this conclusion. The external environments in more traditional accretion disks that launch jets have much lower densities that that of the accretion disks. In CCSNe, on the other hand, the entire volume inner to the stalled shock and close to it has about the same density and there is no high density disk. As well, outside the stalled shock there is a large ram pressure of the in-falling core material. No such ram pressure exists in more traditional cases of disks that launch jets. This might imply that just as the jets start to expand in the SASI zone their environment suppresses their propagation.  Therefore, future studies will have to study the exact mechanism by which the SASI zone launches jets. I do note here that heating by neutrinos that the newly born NS emits can aid the propagation of the jets. I discuss this point in section \ref{sec:Energy}.

\section{energy Considerations}
\label{sec:Energy}

There are two points regarding the energetic of the SASI-driven jets in the frame of the jittering explosion mechanism.

In Keplerian accretion disks the net force on the gas (before we consider the role of magnetic fields) is very small, practically zero, because the centrifugal force balances gravity.
In a sub-Keplerian accretion belt that \cite{SchreierSoker2016} studied the accreted gas reaches such a balance only very close to the surface of the accreting body, a neutron star in the present case, where pressure becomes important.

In the case of SASI-driven jets thermal pressure at the base of the jets (or bipolar outflow) might play a similar role to that of the centrifugal force in Keplerian accretion disks. Below the stalled shock there is a gain region where neutrino heating overcomes neutrino cooling (e.g., \citealt{Mulleretal2017, OConnorCouch2018}).
I argue here that neutrino heating does play a role in the jittering jets explosion mechanism, but in helping the magnetic activity to launch jets and in aiding the jets to locally revive the stalled shock, rather than in globally reviving the stalled shock (as required in the delayed neutrino mechanism).

Numerical simulations show that the stalled shock is very close to being revived by neutrino heating. However, in most numerical simulations the neutrino heating alone does not revive the stalled shock (see section \ref{sec:intro}). I argue here that the jets, or bipolar outflow, that the SASI launches gives the extra energy boost to let some gas to locally break through the stalled shock and expand outward to later explode the star. In other words, the jittering jets locally revive the stalled shock at the momentarily polar directions.

The general process by which neutrino aid jets works as follows. As the jets starts to propagate through the SASI region toward the stalled shock and then out into the in-falling core material, they pass through strong shock waves. Behind the shock waves the very hot jets' material loses energy by neutrino emission. That the shocks occur in the gain region implies that heating by neutrinos coming from the newly born NS compensates for this energy loss. In other words, neutrino heating aids the jittering jets explosion mechanism by reducing the post-shock cooling near the stalled shock.  

The second point concerns the explosion energy.
The velocity amplitude of $A_{\rm SASI} \simeq 2 \times 10^4 \km \s^{-1}$ implies that the available kinetic energy due to rotational velocity  is $\approx 0.5 \Delta  M_{\rm acc} A^2_{\rm SASI}$,
where $\Delta M_{\rm acc}$ is the mass that is accreted during the activity of the spiral SASI.
The energy can be lower, but for the amplification of the magnetic fields the radial velocity also plays a role as it contributes to $\nabla \times {\bf v}$. The kinetic energy due to radial motion is comparable to, and even larger than, that due to azimuthal velocity (e.g., \citealt{Mulleretal2017, OConnorCouch2018}).
I take a fraction of $\eta \simeq 0.5$ of this energy to be transfered to the gas that is ejected in the jets to yield  an explosion energy of
\begin{equation}
\begin{split}
E_{\rm exp} & ({\rm jittering}) \approx 10^{51}
\left( \frac{\eta}{0.5} \right)
\\
&
\times
\left( \frac{\Delta  M_{\rm acc} }{0.5 M_\odot} \right)
\left( \frac{A_{\rm SASI}}{2 \times 10^4 \km \s^{-1}}\right)^2 \erg.
\end{split}
\label{eq:Eexp}
\end{equation}

Expression (\ref{eq:Eexp}) is very crude, and has the following properties.
(1) It is applicable only for the case of jittering jets, and not for jets in super-energetic (super-luminous) CCSNe. In  super-energetic CCSNe the accretion is through a Keplerian accretion disk \citep{Gilkisetal2016} where the efficiency is much larger, by about an order of magnitude per unit accreted mass than what equation (\ref{eq:Eexp}) gives. This is despite that the amplification of the magnetic fields can be as in a Keplerian accretion disk (section \ref{sec:alternating}).
(2) Adding an initial (even low) rotation to the pre-collapse core might increase the efficiency of this mechanism by enlarging the value of $A_{\rm SASI}$.
(3) We can substitute some typical numbers. If SASI starts after a baryonic mass of $1.2 M_\odot$ has been accreted, then an explosion energy of $10^{51}$ crudely requires the formation of a neutron star of a baryonic mass of $\simeq 1.7 M_\odot$, or of a gravitational mass of $\simeq 1.5 M_\odot$.
(4) If the jets are launched from $\simeq 100 \km$ with a terminal velocity of the escape speed from there, $\simeq 6 \times 10^4 \km \s^{-1}$, then the mass in the jets for an explosion energy of $10^{51} \erg$ is $\approx 0.03 M_\odot$, about five per cent of the accreted mass.
(5) The jittering jets explosion mechanism works in a negative feedback mechanism. Once the jets explode the core, accretion stops. Therefore, this mechanism can account also for much weaker explosions, down to $\simeq 10^{50} \erg$.

\section{Summary}
\label{sec:summary}

The failure of the delayed neutrino mechanism to yield a consistent and robust explosion and the observational indications that jets play a significant role in at least some CCSNe (see details in section \ref{sec:intro}), hinted/motivated/forced us to develop the alternative jittering-jets explosion mechanism. The main challenge of the jittering jets explosion mechanism is to lunch jets even when the core material that the newly born neutron star accretes has a sub-Keplerian specific angular momentum. Until now we have assumed that an accretion belt around the newly born neutron star launches the jittering jets (e.g., \citealt{SchreierSoker2016, Soker2019}). Here, for the first time, I incorporated the entire unstable zone of the accretion flow to the jittering jets explosion mechanism. In this unstable zone the spiral SASI modes lead to local non negligible angular momentum of the accreted gas, despite that the average angular momentum is zero.

In section \ref{sec:alternating} I took results from published numerical simulations and presented a new comparison of the shear and vorticity in the spiral SASI zone to those in Keplerian accretion disks. From that comparison I suggested that the alternating shear and local vorticity in this spiral SASI zone can amplify the magnetic fields much as Keplerian accretion disks do. Since Keplerian accretion disks are known to launch jets, I argue that the spiral SASI zone can also launch jets. Although the shear is similar to that in Keplerian accretion disks, the rotational velocity is smaller, such that the available kinetic energy from rotational velocity is smaller, by about an order of magnitude, relative to that in Keplerian accretion disks. In equation (\ref{eq:Eexp}) I very crudely estimated the energy that can be carried by jittering jets for cases where the pre-collapse core does not rotate.

A main point of the newly discussed SASI-driven jets is that neutrino heating plays a significant role in the jittering jets explosion mechanism, but in boosting the launching of jets by magnetic fields and in further energizing the propagation of the jets through the stalled shock, rather than by directly reviving the entire stalled shock. With this neutrino heating and with the magnetic activity (section \ref{sec:alternating}), the accretion of $\approx 0.1-1 M_\odot$ through the spiral SASI and by launching of $\simeq 5-10 \%$  of this mass into jets, the jittering jets explosion mechanism might account for CCSNe with explosion energies of up to
${\rm several} \times 10^{51} \erg$. More energetic supernovae require the formation of a Keplerian accretion disk.
I raise here the possibility that in an intermediate range, where the specific angular momentum of the accreted gas is just below the Keplerian value, an accretion belt does play a role in launching jets \citep{Soker2019}.
 I summarize these three accretion flows to launch jets in Table \ref{table:nonlin}.
\begin{table*}[t]
 \centering \caption{Jet launching cases in CCSNe}
 \begin{tabular}{c c c c} 
 \hline
 Physical parameter      &      Accretion disk          &  Accretion Belt  & Alternating shear \\
                         & \cite{Gilkisetal2016}        &  \cite{Soker2019}& This study  \\
  [0.5ex]
 \hline \hline
Jets' axis    & Constant direction  & Jittering  & jittering  \\
\hline
Pre-collapse  & Rapid & Moderate & Slow to moderate \\
core rotation &       &          & \\
\hline
Average accreted specific        & Continuous           & Continuous/Varying   & Varying    \\
angular momentum ($j_{\rm acc}$) & $\gg j_{\rm Kep}(R)$ & $\la j_{\rm Kep}(R)$ & $\ll j_{\rm Kep}$ \\
 \hline
 Magnetic fields        & Shear in one azimuthal& Shear in one azimuthal  & Local shear zones \\
 amplification          & direction in the disk & direction in inner zone & by alternating shear \\
 \hline
Angular velocity                 &Monotonic          &       Monotonic                     &alternating\\
at launching ($\Omega_ {\rm L}$) &$\Omega_{\rm Kep}$ & ${\rm few}\times0.1\Omega_{\rm Kep}$&${\rm few}\times0.1\Omega_{\rm Kep}$\\
\hline
Vorticity at launching & Monotonic               &  Monotonic           &Alternating \\
($\omega$)             & $(1/2)\Omega_{\rm Kep}$ & $< \Omega_{\rm Kep}$ &$\Omega_{\rm Kep}-{\rm few}\times\Omega_{\rm Kep}$\\
 \hline
 Launching area ($D_{\rm L}$)  &  $\gg R$               &    $\simeq R$ & several$\times R$\\
  \hline
Additional            & Centrifugal      &   Not specified  & Pressure due to \\
Outward force         &                  &                  & neutrino heating\\
 \hline
Transferring accretion& Magnetic fields  &  Magnetic fields   & Magnetic fields    \\
energy to jets        &                  &  +neutrino heating & + neutrino heating \\
 \hline
  \end{tabular}
\label{table:nonlin}
\newline
Three types of jet-launching sites for jets that explode CCSNe via the jet feedback mechanism.
 The different symbols have the following meaning: $R$ is the
 radius of the accreting body; $j_{\rm Kep}$ and $\Omega_{\rm Kep}$ are the Keplerian specific angular momentum and angular velocity, respectively.
\end{table*}

 I call for numerical studies of the explosion of CCSNe to examine the possible implications of their findings to the launching of jets by the spiral SASI inflow.
 Although this is impossible to do directly with presently available numerical codes, I encourage a detail comparison to the properties of Keplerian accretion disks that are known to launch jets. Numerical simulations might add two opposite jets along the momentarily angular momentum axis at any time when there is a developed spiral-SASI that has shear similar to that in Keplerian accretion disks. I predict that the added jittering jets will locally revive the stalled shock and lead to the explosion of the star.

I thank an anonymous referee for helpful comments. This research was supported by the E. and J. Bishop Research Fund at the Technion and by a grant from the Israel Science Foundation.

\end{document}